\documentclass[prl,twocolumn,amsmath,amssymb,aps,showpacs]{revtex4}
\usepackage{graphicx}
\usepackage{bm}
\usepackage{subfigure}
\usepackage{comment}
\usepackage{color}

\usepackage{soul}

\newcommand{\pr}{Phys. Rev.\ }

\newcommand{\etal}{{\em et al.}}
\newcommand{\e}{\mbox{e}}

\newcommand{\UQ}{School of Mathematics and Physics, University of Queensland, Brisbane, 
QLD 4072, Australia.}

\begin{document}

\narrowtext

\title{Non-Gaussian continuous-variable entanglement and steering}

\author{M. K. Olsen and J. F. Corney}

\affiliation{\UQ}
\date{\today}

\begin{abstract}

Two Kerr-squeezed optical beams can be combined in a beamsplitter to produce non-Gaussian continuous-variable entangled states. We characterize the non-Gaussian nature of the output by calculating the third-order cumulant of quadrature variables, and predict the level of entanglement that could be generated by evaluating the Duan-Simon and Reid Einstein-Podolsky-Rosen criteria.  These states have the advantage over Gaussian states and non-Gaussian measurement schemes in that the well known, efficient and proven technology of homodyne detection may be used for their characterisation. A physical demonstration maintaining the important features of the model could be realised using two optical fibres, beamsplitters, and homodyne detection.

\end{abstract}

\pacs{42.50.Dv, 42.50.Lc, 03.67.Bg, 03.67.Mn}

\maketitle

\section{Introduction}

Continuous variable (CV) systems provide flexible and powerful means for implementing quantum information schemes~\cite{Braunstein}, in large part because there are mature and precise techniques for measuring the quadratures of light, most of which are familiar from classical communications technologies. Despite the need to deal with transmission losses, recent work has demonstrated useful distances  comparable to those achieved with discrete-variable systems~\cite{Grangier}.  Furthermore,  research and development has progressed to the stage where CV quantum key distributions systems have advantages over discrete-variable methods~\cite{Grosshans,Lance,Lam}.

The remaining stumbling block to the wider use of CV systems is that the most readily available CV systems and the the most developed detection techniques produce only Gaussian statistics.  This limitation rules out tasks such as entanglement distillation~\cite{Plenio}, quantum error correction~\cite{Niset}, and quantum computation.  One way of introducing non-Gaussian statistics is through nonlinear measurements ~\cite{Lvovsky}, but this approach negates one of the main advantages of CV systems, namely, the highly developed technology that is available for performing Gaussian homodyne measurements. 

In this paper, we proceed along an alternative approach, namely to use CV sources that produce non-Gaussian outputs.  The importance of this area of research was shown recently by Ohliger \etal~\cite{Eisert}, who demonstrated there are serious limitations to the use of Gaussian states for quantum information tasks which may be avoided by developing useful and relatively simple non-Gaussian sources.  


Non-Gaussian light can be produced by means of a $\chi^{(3)}$ nonlinear medium, such as a single-mode optical fibre. Pairs of such Kerr squeezed beams can be combined on a beam splitter to produce CV entangled states, as has been experimentally produced using polarisation squeezing~\cite{Dong}.  However, the non-Gaussian character of these states has not to our knowledge been explicitly demonstrated.

In this paper, we use a single-mode anharmonic oscillator~\cite{Gerard} to determine the non-Gaussian entanglement that can in principle be achieved with Kerr-squeezed states.   We characterise the non-Gaussian statistics through higher-order cumulants and gauge the level of entanglement by calculating the Duan-Simon and Einstein-Podolsky-Rosen correlations. 

\section{Testing for non-Gaussian statistics}

A Gaussian state can be most simply defined as a state with a Gaussian Wigner function, i.e. a state whose marginal distributions are Gaussian.  For a CV state, the departures from non-Gaussian behaviour can thus be characterised by the skewness of the distributions of its quadrature moments, as revealed in nonzero higher-order cumulants \cite{Dubost}.



We define the generalised quadrature $\hat{X}(\theta)$ at quadrature at angle $\theta$ as 
\begin{equation}
\hat{X}(\theta) = \hat{a}\e^{-i\theta}+\hat{a}^{\dag}\e^{i\theta},
\label{eq:quads}
\end{equation}
so that the canonical $\hat{X}$ quadrature is found at $\theta=0$, with conjugate
$\hat{Y}=\hat{X}(\frac{\pi}{2})$.  

For a Gaussian distribution, all cumulants higher than second order vanish, and therefore we can test for non-Gaussian statistics by a the presence of a nonzero third-order cumulant:
\begin{equation}
\kappa_3(\theta) = \langle \hat{X}^{3}(\theta)\rangle+2\langle\hat{X}(\theta)\rangle^{3}-3\langle\hat{X}(\theta)\rangle\langle\hat{X}^{2}(\theta)\rangle.
\label{eq:skewness3}
\end{equation}
While $\kappa_3 \neq 0$ is a sufficient condition for non-Gaussian statistics, it is not a necessary one.  In particular, $\kappa_3$ will vanish for a symmetric distribution in phase space.  In the presence of such symmetry, the fourth-order moment $\kappa_4$ provides the lowest-order test for non-Gaussian behaviour:
\begin{equation}
\kappa_4(\theta) =   \langle \hat X^4(\theta) \rangle + 2\langle \hat X(\theta) \rangle^4 - 3\langle \hat X^2(\theta) \rangle^2 - \langle \hat X(\theta) \rangle \kappa_3(\theta).
\end{equation}
\label{eq:skewness4}

The fourth-order cumulant can be used to infer the negativity of the Wigner function \cite{Bednorz}, which is considered to be a direct measure of the nonclassicality of a state.  It also allows comparison to the nonclassical states that have been experimentally demonstrated to be non-Gaussian, such as the number state \cite{Lvovsky} and the photon-subtracted squeezed vacuum \cite{Wenger,Kim}.  For both of these states, $\kappa_4$ scales quadratically with number.  For the number-state for example,
\begin{equation}
\kappa_4 = -6n(n+1).
\end{equation} 
In the analysis below, we will determine the regimes in which the Kerr-squeezed state is skewed to a similar level.

\section{Non-Gaussian statistics in the Kerr-squeezed state}

The Hamiltonian for the single-mode model, ignoring any effects due to loss and excess noise, is
\begin{equation}
{\cal H}=\hbar\chi\left(a^{\dag}a\right)^{2},
\label{eq:kerr1}
\end{equation}
where $\chi$ represents the third-order nonlinearity of the medium and $\hat{a}$ is the bosonic annihilation operator for the electromagnetic field mode.

For an input Glauber-Sudarshan coherent state,
\begin{equation}
|\alpha\rangle = \e^{-|\alpha|^{2}/2}\sum_{n=0}^{\infty}\frac{\alpha^{n}}{\sqrt{n!}}|n\rangle,
\label{eq:coherent}
\end{equation}
where $|n\rangle$ represents a Fock state of fixed number, we may find analytical expressions for all the operator moments necessary to calculate the first four cumulants. 

The Heisenberg equation of motion for $\hat{a}$ can formally be solved to give
\begin{equation}
\hat{a}(t) = \e^{-i\chi t(2\hat{a}^{\dag}\hat{a}+1)}\hat{a}(0),
\label{eq:at}
\end{equation}
whose expectation value in a coherent state is
\begin{eqnarray}
\langle \hat{a}(t)\rangle &=& \alpha\e^{-i\chi t}\e^{|\alpha|^{2}(\cos 2\chi t - i\sin 2\chi t -1)}.
\label{eq:coherentexp1}
\end{eqnarray}


Defining $\hat a_\theta \equiv \hat a e^{-i\theta}$, we can write the mean of a quadrature moment as
\begin{equation}
\langle\hat X(\theta,t)\rangle =\langle \hat{a}_\theta(t)+\hat{a}_\theta^{\dag}(t)\rangle,
\label{eq:Xmean}
\end{equation}
for which we already have a solution. The second moment is
\begin{equation}
\langle \hat{X}^{2}(\theta,t)\rangle = \langle 1+ 2\hat{a}_\theta^{\dag}\hat{a}_\theta+\hat{a}_\theta^{\dag\,2}+\hat{a}_\theta^{2}\rangle,
\label{eq:Xsq}
\end{equation}
where we have dropped the time argument on the RHS for simplicity. The third- and fourth-order moments are
\begin{eqnarray}
\langle \hat{X}^{3}(\theta)\rangle &=&\langle \hat{a}_\theta^{\dag\,3}+3\hat{a}_\theta^{\dag\,2}\hat{a}_\theta+3\hat{a}_\theta^{\dag}\hat{a}_\theta^{2} +\hat{a}_\theta^{3} +3\hat{a}_\theta^{\dag}+3\hat{a}_\theta\rangle, \nonumber \\
\langle \hat{X}^{4}(\theta)\rangle &=& \langle a_\theta^4 + 4a_\theta^\dagger a_\theta^3 + 6{a_\theta^\dagger}^2a_\theta^2 + 4 {a_\theta^\dagger}^3a_\theta + {a_\theta^\dagger}^4 \nonumber \\
&& + \, 6a_\theta^2 + 12a_\theta^\dagger a_\theta + 6{a_\theta^\dagger}^2+ 3 \rangle.
\label{eq:Xcube}
\end{eqnarray}
To analytically calculate these moments, we use the following expectation values and their complex conjugates:
\begin{eqnarray}
%
\langle \hat{a}_\theta^{2}(t)\rangle &=& \alpha^{2}e^{-2i\theta} \e^{-4i\chi t}\e^{|\alpha|^{2}(\cos 4\chi t-i\sin 4\chi t-1)},\nonumber\\
%
%
\langle \hat{a}_\theta^{3}(t)\rangle &=& \alpha^{3}e^{-3i\theta}\e^{-9i\chi t}\e^{|\alpha|^{2}(\cos 6\chi t-i\sin 6\chi t-1)},\nonumber\\
%
%
\langle \hat{a}_\theta^{\dag}\hat{a}_\theta^{2}(t)\rangle &=& \alpha^{\ast}\alpha^{2}e^{-i\theta}\e^{-3i\chi t}\e^{|\alpha|^{2}(\cos 2\chi t-i\sin 2\chi t-1)}, \nonumber \\
%
\langle a_\theta^4 (t)\rangle &=& \alpha^4e^{-4i\theta} e^{-i 16\chi t} e^{|\alpha|^2 (\cos{8\chi t} - i\sin{8\chi t} - 1)}, \nonumber \\
\langle a_\theta^\dagger a_\theta^3 (t)\rangle &=& \alpha^*\alpha^3 e^{-2i\theta} e^{-i 8\chi t}e^{|\alpha|^2 (\cos{4\chi t} -i \sin{4\chi t}- 1)}.\nonumber\\
\label{eq:moments}
\end{eqnarray}

These equations reveal several kinds of contributions to the dynamics, with different time scales.  The sin and cos terms in the exponents can each be expanded, and for sufficiently small interaction time $\chi t$, we can keep the first two terms in each, i.e. up to fourth order in time.  We are left with a number of different contributions to the exponent.

First, there is the nonlinear phase factor  proportional to $N \chi t$, where $N = |\alpha|^2$.  This mean-field frequency shift  can be removed by a switch to a rotating frame, i.e. setting $\theta = \theta_0 + 2 N\chi t$.  

Second, the real exponent proportional to $N\chi^2t^2$ is responsible for squeezing, although in order to obtain {\em quantum} squeezing, i.e. below the coherent-state level, we also require the zero-point phase factors.

Finally, there are the third and fourth order terms $N\chi^3t^3$ and $N\chi^4t^4$, which for large N give the leading order contribution to the third and fourth order cumulants and hence are responsible for most of the skewness we see in the quadrature statistics. 

In typical Kerr squeezing experiments, the number of photons is large $N\gg 1$ in order to compensate for a weak nonlinearity $\chi \ll 1$.  In this limit, we can derive a simple expression for the third-order cumulant of the $Y$ quadrature rotating at the mean-field frequency, which is where skewness is most evident:

\begin{equation}
\kappa_3\left(\frac{\pi}{2}\right) \simeq -256 \frac{1}{\sqrt{N}} \left(\chi N t\right)^3. \label{simple}
\end{equation}

\begin{figure}[tbhp]
\begin{center} 
\includegraphics[width=\columnwidth ]{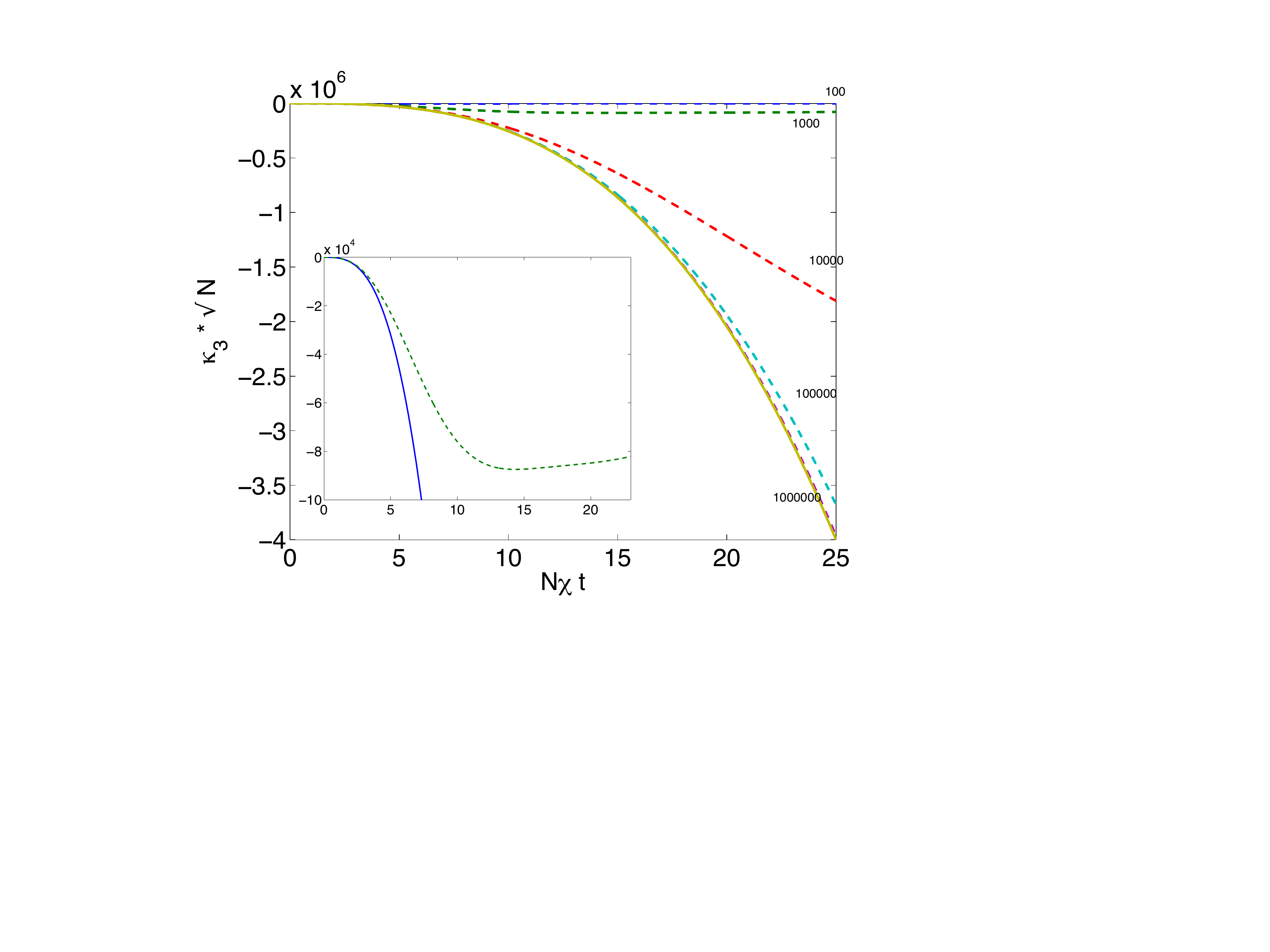}

\end{center}  
\caption{(colour online) The third order cumulant, $\kappa_{3}$, of  $\hat Y = \hat X(\pi/2)$ in a rotating frame as a function of time, for various photon numbers ranging from 100 to $10^6$, as labeled.  In this and subsequent plots, time is scaled by the inverse of the mean-field interaction strength and hence is a dimensionless quantity; $\kappa_3$ is scaled by $1/\sqrt{N}$.  The dashed lines give the exact results (Eqs.~(\ref{eq:moments})) and the solid line gives the approximate result (Eq.\ (\ref{simple})), which is accurate for large $N$ or small mean-field interaction time $N\chi t$. The inset shows $\kappa_3$ for $N=1000$ in more detail.}
\label{fig:skew3}
\end{figure}

\begin{figure}[tbhp]
\begin{center} 
\includegraphics[width=\columnwidth ]{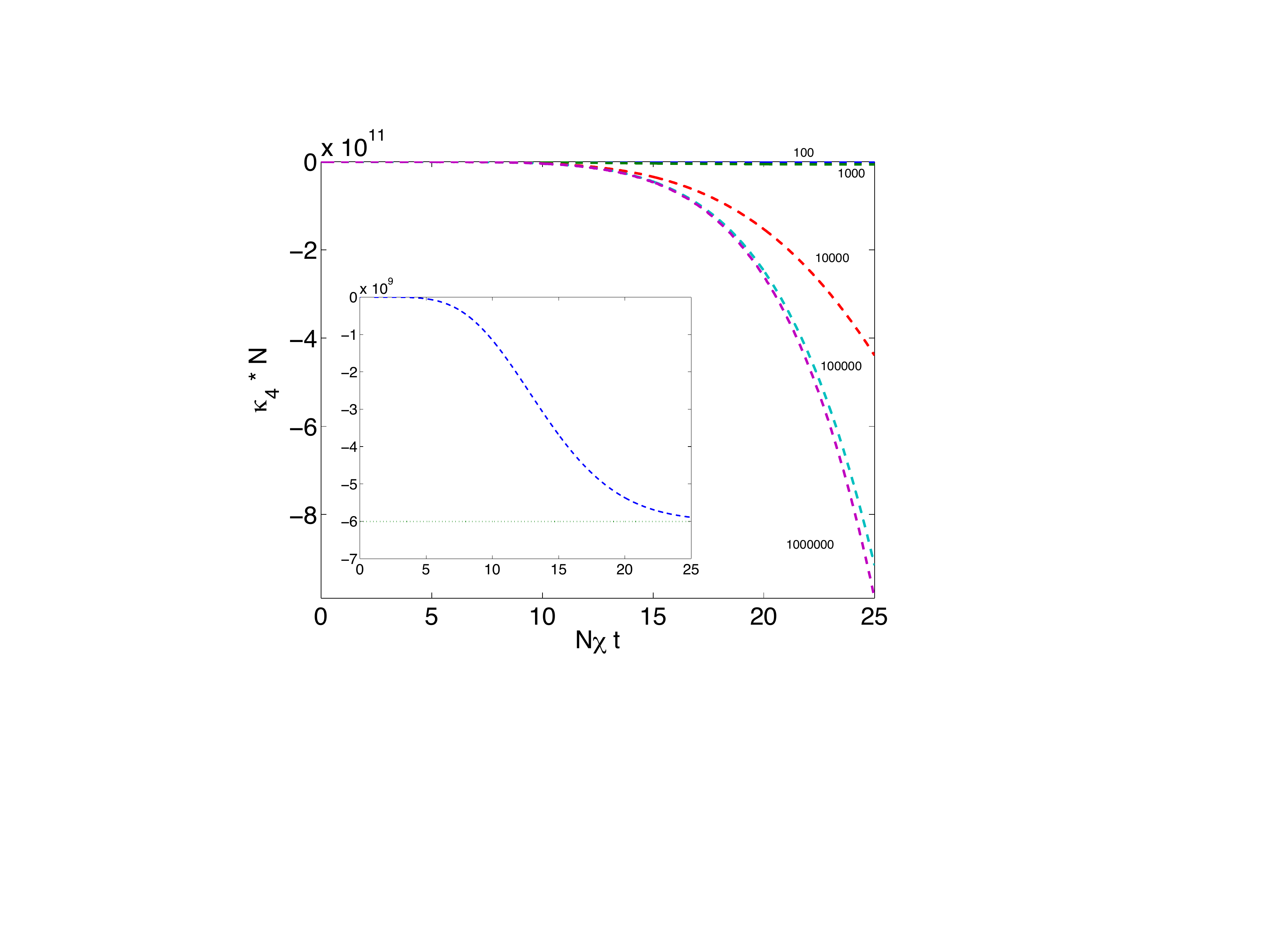}

\end{center}  
\caption{(colour online) The fourth-order cumulant, $\kappa_{4}$ of $\hat Y = \hat X(\pi/2)$ in a rotating frame as a function of time, for various photon numbers ranging from 100 to $10^6$, as labeled.  Time is scaled by the mean-field interaction strength, and $\kappa_3$ is scaled by $1/{N}$.  For large $N$ or small mean-field interaction time $N\chi t$, the results approach the same limiting curve $\propto t^4$. The inset shows $\kappa_4$ for $N=1000$ in more detail, both for the Kerr-squeezed state (dashed line) and for the number state (dotted line).}
\label{fig:skew4}
\end{figure}

The validity of this expression is demonstrated in Fig.~\ref{fig:skew3}, which plots the third-order cumulant for various photon numbers.  The exact results for $N > 10^6$ are indistinguishable on this time scale from the simple cubic  growth described by Eq.~(\ref{simple}).  Time is here scaled by $N\chi$ in order compare results that give the same Kerr effect.  On this scale, the third-order cumulant decreases in proportion to the square root of the number of photons.   Note however, that the absolute size of the third-order cumulant increases with particle number, at a rate faster than $\langle \hat X \rangle^3$:    
\begin{equation}
\frac{\kappa_3}{\langle \hat X \rangle^3} \sim N .
\end{equation}

The fourth-order cumulant $\kappa_4$ is plotted in  Fig.~\ref{fig:skew4}.  Again, for large photon numbers the cumulant approaches a limiting scaling behaviour: 
\begin{equation}
\kappa_4 \propto \frac{1}{N} (\chi N t)^4,
\label{simple4}
\end{equation}
which gives the same relative growth of 
\begin{equation}
\frac{\kappa_4}{\langle \hat X \rangle^4} \sim N, 
\end{equation}
although if the time is adjusted as a function of $N$ to keep the Kerr squeezing constant, the fourth-order cumulant decreases in proportion to the particle number.

\begin{figure}[htbp] 
   \centering
   \includegraphics[width=\columnwidth,trim=0cm 6cm 1cm 6cm, clip]{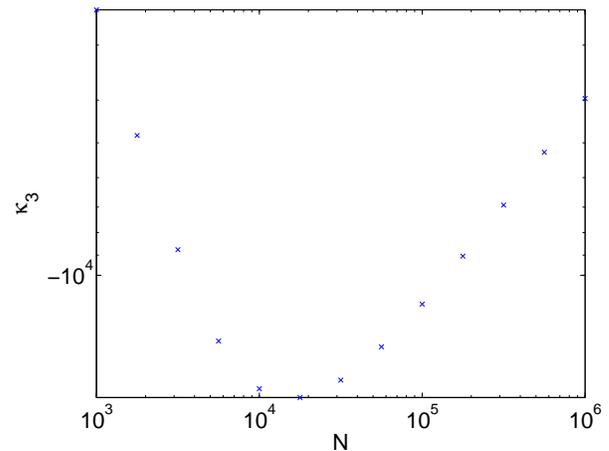} 
   \caption{(colour online) Third-order cumulant $\kappa_3(\frac{\pi}{2})$ of the Kerr-squeezed state at fixed $\chi N t = 25$ as a function of $N$.   The behaviour at $N\gtrsim10^5$ reveals the scaling behaviour described by Eq.\ (\ref{simple}); the behaviour at $N\lesssim10^4$ is due to the saturation effect seen in the inset to Fig.\ \ref{fig:skew3}.}
   \label{fig:exactN3}
\end{figure}

\begin{figure}[htbp] 
   \centering
      \includegraphics[width=\columnwidth,trim=0cm 6cm 1cm 6cm,clip]{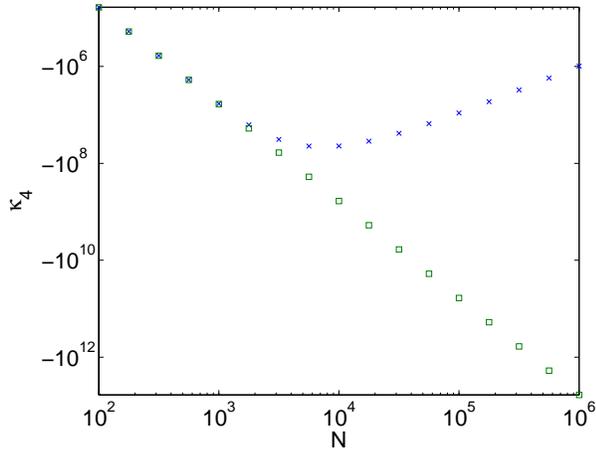} 
   \caption{(colour online) Fourth-order cumulant $\kappa_4(\frac{\pi}{2})$ of the Kerr-squeezed state at fixed $\chi N t = 25$ as a function of $N$ (crosses).  The squares give the fourth-order cumulant of a number state with the same number of photons.  For $N \lesssim10^3 $
 the Kerr-squeezed result saturates to that of the number state (see inset to Fig.\ \ref{fig:skew4}). For $N \gtrsim10^4$, the results follow the scaling as given in Eq.~(\ref{simple4}).}
    \label{fig:exactN4} 
\end{figure}

Figures  \ref{fig:exactN3} and \ref{fig:exactN4} show the cumulants as a function of $N$, for a fixed value of $\chi N t = 25$.  One can clearly see two different regimes of behaviour, with the crossover between the two occurring just above  $N \sim 10^4$.  For $\kappa_4$ the number-state result is also plotted for comparison. One can see that $\kappa_4/(Nt)^4$ scales as described above for large $N$, but for small $N$ is limited to values of the order of corresponding number-state results (increasing with $N$ quadratically).  This result suggests that a Kerr-squeezed state can be as non-Gaussian by this measure as the number state, for sufficiently long interaction time.



\section{Quadrature variances, entanglement and  Einstein-Podolsky-Rosen correlations}

\begin{figure}[tbhp]
\begin{center} 
\includegraphics[width=\columnwidth,  trim = 1cm 6cm 1cm 9cm, clip]{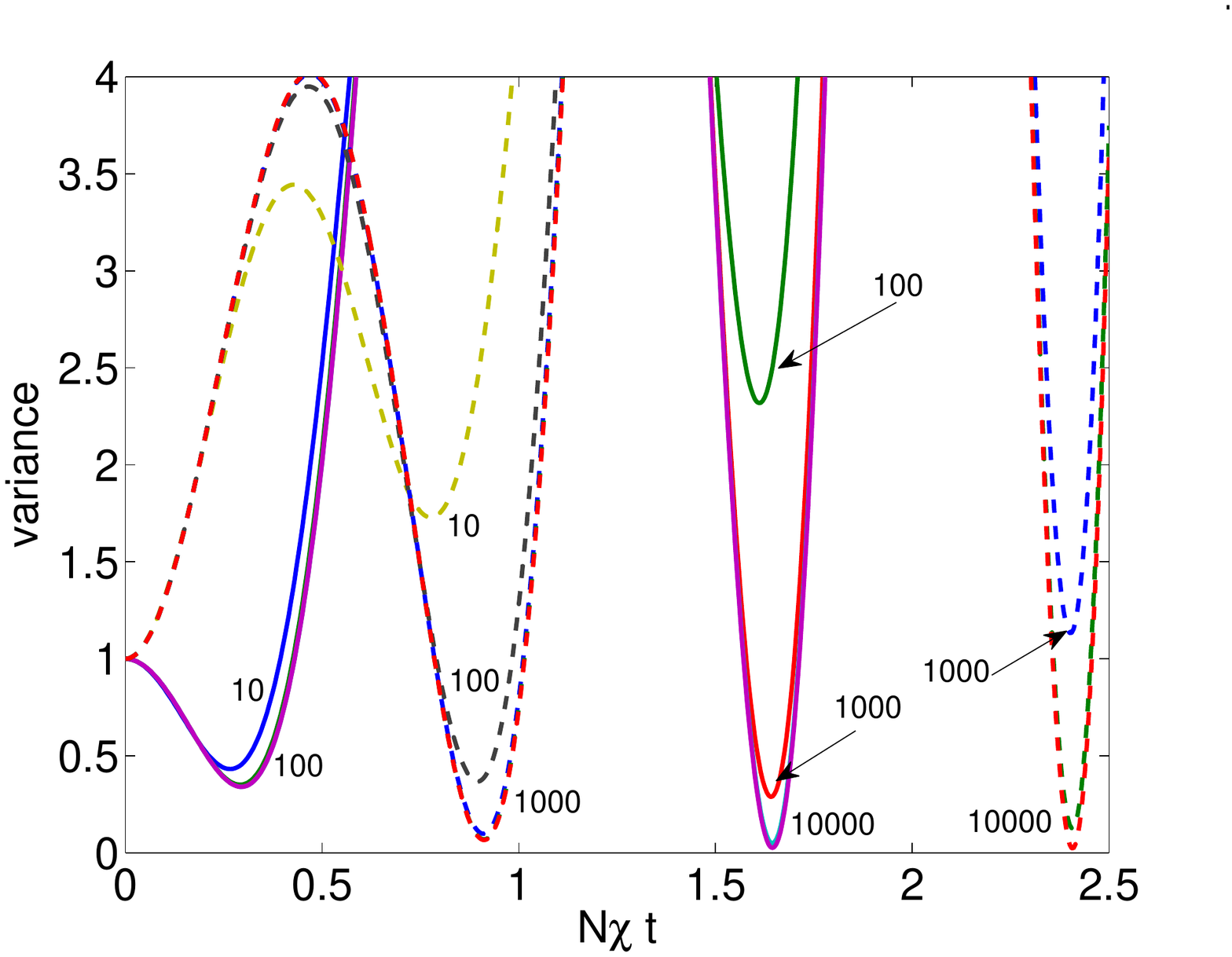}
\end{center}  
\caption{(colour online) The variances in the $\hat{X}$ (solid line) and $\hat{Y}$ (dashed line) quadratures as a function of mean-field interaction time $N\chi t$. On this scale, the squeezing results for different numbers are indistinguishable for $N>10^4$.}
\label{fig:Vshort}
\end{figure}

Although entangled states have already been predicted to be produced by the intracavity nonlinear coupler~\cite{NLC,volante}, the linearisation process used to obtain the spectra in those cases forces Gaussian statistics on the outputs. Here, we wish to produce entangled states that maintain non-Gaussian statistics, so we will proceed by mixing the outputs of two Kerr oscillators on a beamsplitter~\cite{vLB}, as experimentally demonstrated by \cite{Dong}.
We will now show the quadrature variances, as we need squeezed states in order to obtain entangled modes in the outputs. We note that, while this could be done by mixing one squeezed mode with vacuum, better results in terms of the degree of violation of the relevant inequalities are obtained by mixing two squeezed states.
The Heisenberg uncertainty principle demands that
\begin{equation}
V\left(\hat{X}(\theta)\right)V\left(\hat{X}(\theta+\frac{\pi}{2})\right)\geq 1,
\label{eq:HUP}
\end{equation}
so that any quadrature with variance below one is squeezed.
Fig.~\ref{fig:Vshort} shows the variances of the canonical $\hat{X} = \hat X(0)$ and $\hat{Y}= \hat X(\frac{\pi}{2})$ quadratures with time again scaled by $N\chi t$ so that, apart from small-number effects, the same level of squeezing is obtained for different photon numbers.  As for the cumulants, the results for $N>10^4$ cannot be distinguished on this time scale. In fact, above $N = 1000$, the small-number effects only appear as small differences amount of squeezing, so for the remainder of the paper, we quote results for $N=1000$.


Considering a beamsplitter with reflectivity $\eta$ and labelling the inputs by $\hat{a}_{1}$ and $\hat{a}_{2}$ and the outputs by $\hat{b}_{1}$ and $\hat{b}_{2}$,  we find
\begin{eqnarray}
\hat{b}_{1} &=& \sqrt{\eta}\hat{a}_{1}+i\sqrt{1-\eta}\hat{a}_{2},\nonumber\\
\hat{b}_{2} &=& i\sqrt{1-\eta}\hat{a}_{1}+\sqrt{\eta}\hat{a}_{2}.
\label{eq:bsrelation}
\end{eqnarray}
For notational convenience, we will now make the simplification $\hat{X}_{b_{j}}\rightarrow \hat{X}_{j}$, and similarly for $\hat{Y}_{b_{j}}$. This allows us to define the variances of the beamsplitter outputs as, 
\begin{eqnarray}
V(\hat{X}_{1}) &=&  \eta V(\hat{X}_{a_{1}}) + (1-\eta) V(\hat{Y}_{a_{2}}),\nonumber\\
V(\hat{X}_{2}) &=& (1-\eta) V(\hat{Y}_{a_{1}})+\eta V(\hat{X}_{a_{2}}),
\nonumber\\
V(\hat{Y}_{1}) &=& \eta V(\hat{Y}_{a_{1}})+(1-\eta) V(\hat{X}_{a_{2}}),
\nonumber\\
V(\hat{Y}_{2}) &=& (1-\eta)V(\hat{X}_{a_{1}})+\eta V(\hat{Y}_{a_{2}}).
\label{eq:Varbmodes}
\end{eqnarray}

\begin{figure}[tbhp]
\begin{center} 
\includegraphics[width=\columnwidth]{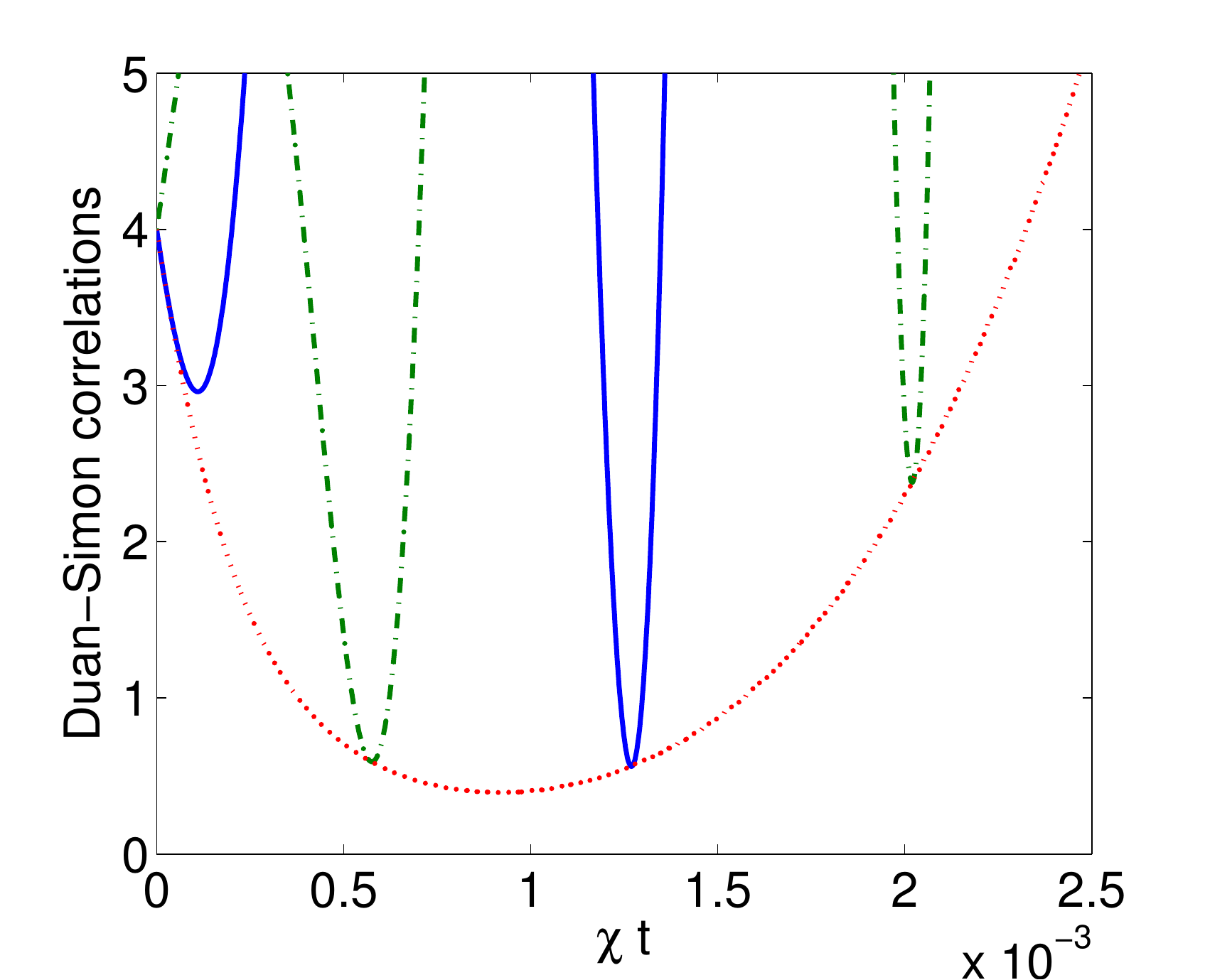}
\end{center}  
\caption{(colour online) Duan-Simon correlations (LHS of Eq.~(\ref{eq:DSinequality})) after mixing on a $50:50$ beamsplitter, as a function of interaction time $\chi t$, for N=1000. A value below $4$ signifies entanglement. The (blue) solid line uses the canonical $\hat{X}_{1}-\hat{X}_{2}$ and $\hat{Y}_{1}+\hat{Y}_{2}$ quadratures and the green (dash-dotted) line uses  the canonical $\hat{X}_{1}+\hat{X}_{2}$ and $\hat{Y}_{1}-\hat{Y}_{2}$ quadratures, while the (red) dotted line is optimised for quadrature angle at each time.}
\label{fig:DSXY}
\end{figure}

Along with the covariances,
\begin{eqnarray}
V(\hat{X}_{1},\hat{X}_{2}) &=& -\sqrt{\eta(1-\eta)}\left[V(\hat{X}_{a_{1}},\hat{Y}_{a_{1}})+V(\hat{X}_{a_{2}},\hat{Y}_{a_{2}})\right],\nonumber\\
V(\hat{Y}_{1},\hat{Y}_{2}) &=& \sqrt{\eta(1-\eta)}\left[V(\hat{X}_{a_{1}},\hat{Y}_{a_{1}})+V(\hat{X}_{a_{2}},\hat{Y}_{a_{2}})\right],\nonumber\\
\label{eq:covars}
\end{eqnarray}
we now have all the expressions needed to calculate the quantities necessary to check for violation of the continuous-variable Duan-Simon inequality~\cite{Duan,Simon}. For the purposes of this article, we define this as 
\begin{equation}
V(\hat{X}_{1}\pm\hat{X}_{2})+V(\hat{Y}_{1}\mp\hat{Y}_{2})\geq 4,
\label{eq:DSinequality}
\end{equation}
with any violation of this inequality being sufficient to demonstrate the presence of entanglement for a non-Gaussian state. The result for this correlation, with $\eta=0.5$, is shown in Fig.~\ref{fig:DSXY}. The (red) dotted line gives the maximum violation, optimised over quadrature angle $\theta$.  Clearly the outputs from two Kerr oscillators mixed on a beamsplitter can give a continuous-variable non-Gaussian entangled bipartite resource. 

\begin{figure}[tbhp]
\begin{center} 
\includegraphics[width=\columnwidth]{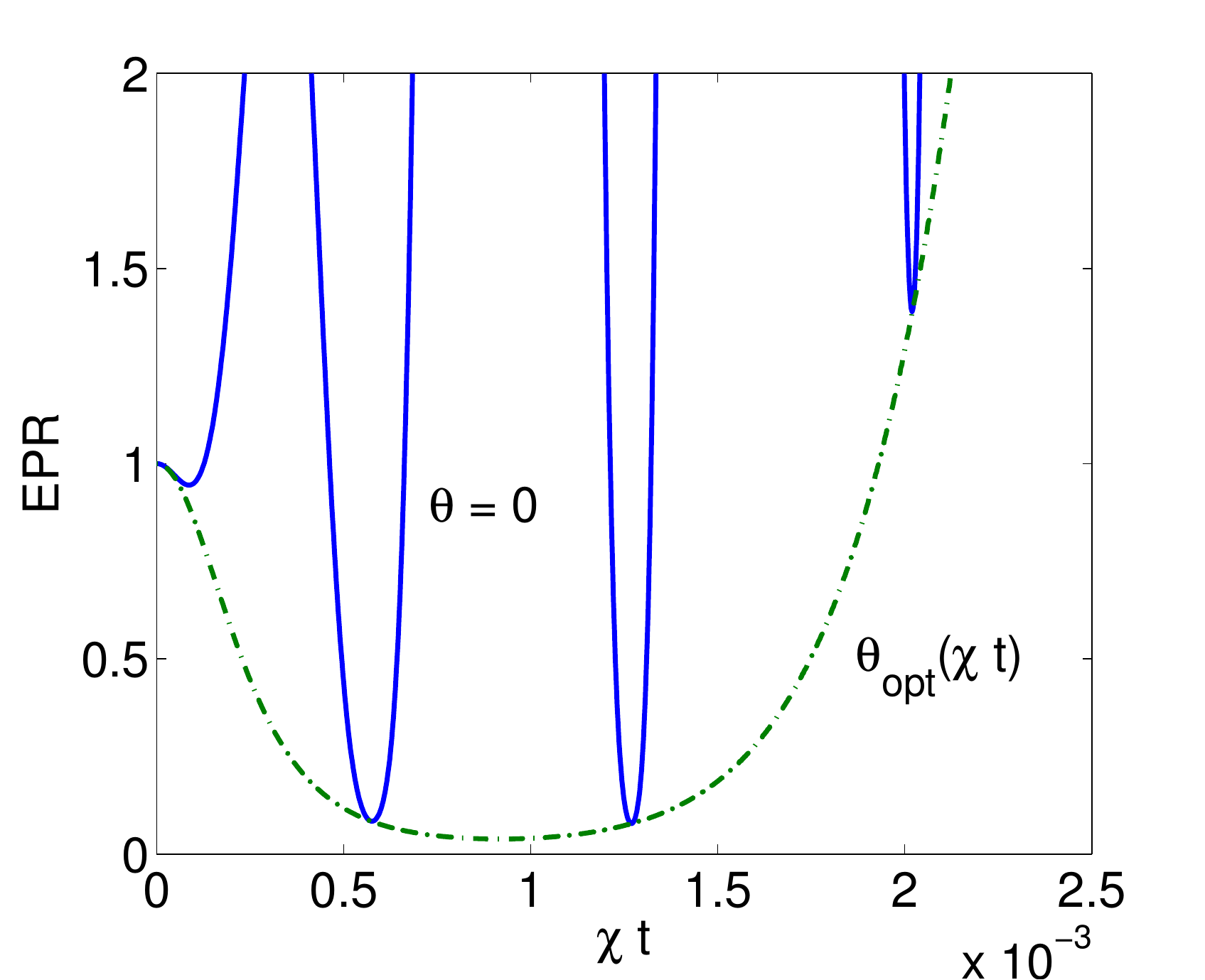}
\end{center}  
\caption{(colour online) Reid EPR correlation (LHS of Eq.~(\ref{eq:EPR})) after mixing on a $50:50$ beamsplitter as a function of interaction time $\chi t$, for $N=1000$. A value below $1$ signifies a demonstration of the EPR paradox. The dashed, lower line is optimised for quadrature angle, while the upper, solid line is for the canonical quadratures.}
\label{fig:EPR}
\end{figure}

As shown by Wiseman \etal~\cite{Howard} and Cavalcanti \etal~\cite{corno}, the inseparability of the system density matrix describes a set of states which includes within it subsets which are more deeply non-classical than evidenced by entanglement alone, such as those which demonstrate the Einstein-Podolsky-Rosen (EPR) paradox~\cite{EPR}. For our purposes here, we will use the inequality developed by Reid~\cite{Margaret}, written as 
\begin{equation}
V_{inf}(\hat{X}_{b_{j}})V_{inf}(\hat{Y}_{b_{j}})\geq 1,
\label{eq:EPR}
\end{equation}
where $j=1,2$ and
\begin{eqnarray}
V_{inf}(\hat{X}_{b_{j}}) &=& V(\hat{X}_{b_{j}})-\frac{\left[V(\hat{X}_{b_{j}},\hat{X}_{b_{k}})\right]^{2}}{V(\hat{X}_{b_{k}})},\nonumber\\
V_{inf}(\hat{Y}_{b_{j}}) &=& V(\hat{Y}_{b_{j}})-\frac{\left[V(\hat{Y}_{b_{j}},\hat{Y}_{b_{k}})\right]^{2}}{V(\hat{Y}_{b_{k}})}.
\label{eq:ReidEPR}
\end{eqnarray}
From the expressions given above for the Duan-Simon criterion, it can be seen that all the moments necessary to calculate these expressions are available analytically.  As shown in Fig.~\ref{fig:EPR}, the two modes after the beamsplitter exhibit a strong violation of the Reid inequality.

Finally, we consider the skewness of the final entangled state. For a $50:50$ beam splitter, the third- and fourth-order cumulants of the $X$-quadrature in output port 1  can be shown to be:
\begin{eqnarray}
\kappa_3(X_1) &=& \frac{1}{\sqrt{8}}\left( \kappa_3(X_{a1}) - \kappa_3(Y_{a2})\right) \nonumber\\
\kappa_4(X_1) &=& \frac{1}{4}\left( \kappa_4(X_{a1}) + \kappa_4(Y_{a2})\right) + 4 \langle X_1\rangle \kappa_3 (X_1),  \nonumber\\
\end{eqnarray}
which confirms that skewed inputs to a beamsplitter lead to skewed outputs, with cumulants generally of the same order of magnitude.

\section{Conclusions}

In summary, we have employed a simple model of $\chi^{(3)}$ nonlinear process to demonstrate that violations of the Duan-Simon and Reid entanglement criteria occur at the same time as significant departures from Gaussian behaviour.  For sufficiently long interaction time $N\chi t$, the nonlinear interaction will skew the distribution of the quadrature variables, leading to large third and fourth order cumulants.  

Moreover, such non-Gaussian entanglement occurs in regimes accessible to optical fibre experiments \cite{Joel1}. However, for accurate quantitative predictions, one would need to go beyond the single-mode model, to include the effects of pulse dynamics and extra noise sources, using the simulations methods, for example, that were used in \cite{Joel2}.

As with any coherent scheme employing the $\chi^{(3)}$ nonlinearity of optical fibres, the intrinsic weakness of the nonlinearity itself can be a limiting factor.  Besides using large photon-numbers, this factor may be overcome by use of electromagnetically induced transparency to produce giant cross-Kerr nonlinear phase shifts and hence highly non-Gaussian quantum states \cite{Tyc}.   On the other-hand, the presence of large numbers of photons makes optical fibre a very bright source of non-Gaussian entanglement, which may well be a practical advantage over number-state schemes.

\section*{Acknowledgments}

MKO gratefully acknowledges support by the Australian Research Council under the Future Fellowships scheme.


\begin{thebibliography}{99}

%
%
%
%
\bibitem{Braunstein}{S.L. Braunstein and P. van Loock, Rev. Mod. Phys. {\bf 77}, 513 (2005).}
%
\bibitem{Grangier}{A. Leverrier and P. Grangier, \prl {\bf 102}, 180504 (2009).}
%
\bibitem{Grosshans}{F. Grosshans, G. Van Assche, J. Wenger, R. Brouri, N.J. Cerf, and P. Grangier, Nature {\bf 421}, 238 (2003).}
%
\bibitem{Lance}{A.M. Lance, T. Symul, V. Sharma, C. Weedbrook, T.C. Ralph, and P.K. Lam, \prl {\bf 95}, 180503 (2005).}
%
\bibitem{Lam}{T.C. Ralph and P.K. Lam, Nature Photonics {\bf 3}, 671 (2009).}
%
\bibitem{Plenio}{J. Eisert, S. Scheel, and M.B. Plenio, \prl {\bf 89}, 137903 (2002).}
%
\bibitem{Niset}{J. Niset, J. Fiur\'a\u sek, and N.J. Cerf, \prl {\bf 102}, 120501 (2009).}
%
\bibitem{Lvovsky}{A.I. Lvovsky, H. Hansen, T. Aichele, O. Benson, J. Mlynek, and S. Schiller, \prl {\bf 87}, 050402 (2001).}
%
\bibitem{Eisert}{M. Ohliger, K. Kieling, and J. Eisert, \pra {\bf 82}, 042336 (2010).}
%
\bibitem{Dong}{R. Dong, J. Heersink, J.-I. Yoshikawa, O. Gloeckl, U. L. Andersen and G. Leuchs,  New J. Phys.\  {\bf 9}, 410 (2007).}
\bibitem{Gerard}{G.J. Milburn, \pra {\bf 33}, 674 (1985).}
%
\bibitem{Dubost}{B. Dubost, M. Koschorreck, M. Napolitano, N. Behbood, R. Sewell and M. Mitchell, \prl {\bf 108}, 183602 (2012).}
%
\bibitem{Bednorz}{A. Bednorz and W. Belzig, \pra  {\bf 83}, 052113 (2011).}
%
\bibitem{Wenger}{J. Wenger, R. Tualle-Brouri and Philippe Grangier, \prl {\bf 92}, 153601 (2004).}
%
\bibitem{Kim}{M. S. Kim,  E. Park,  P. L. Knight, and H. Jeong, \pra{\bf 71}, 043805 (2005).}
%
\bibitem{NLC}{M.K. Olsen, \pra {\bf 73}, 053806 (2006).}
%
\bibitem{volante}{S.L.W. Midgley, A.J. Ferris, and M.K. Olsen, \pra {\bf 81}, 022101 (2010).}
%
\bibitem{vLB}{P. van Loock and S.L. Braunstein, \prl {\bf 84}, 3482 (2000).}
%
%
\bibitem{Duan}{L.-M. Duan, G. Giedke, J.I. Cirac, and P. Zoller, \prl {\bf 84}, 2722 (2000).}
%
\bibitem{Simon}{R. Simon, \prl {\bf 84}, 2726 (2000).}
%
\bibitem{Howard}{H.M. Wiseman, S.J. Jones, and A.C. Doherty, \prl {\bf 98}, 140402 (2007).}
%
\bibitem{corno}{E. G. Cavalcanti, S. J. Jones, H. M. Wiseman, and M. D. Reid,  \pra {\bf 80}, 032112 (2009).} 
%
\bibitem{EPR}{A. Einstein, B. Podolsky, and N. Rosen, \pr {\bf 47}, 777 (1935).}
%
\bibitem{Margaret}{M.D. Reid, \pra {\bf 40}, 913 (1989).}
%
\bibitem{Joel1}{R. Dong, J. Heersink, J.F. Corney, P.D. Drummond, U.L. Andersen, and G. Leuchs, \ol {\bf 33}, 116 (2008).}
%
\bibitem{Joel2}{J.F. Corney, J. Heersink, R. Dong, V. Josse, P.D. Drummond, G. Leuchs, and U.L. Andersen, \pra {\bf 78}, 023831 (2008).}
%
\bibitem{Tyc}{T. Tyc and N. Korolkova, New J. Phys. {\bf 10}, 023401 (2008).}
%
\end{thebibliography}
\end{document}